\documentclass[conference, letterpaper]{IEEEtran}
\IEEEoverridecommandlockouts

\usepackage{cite}
\usepackage{amsmath,amssymb,amsfonts}
\usepackage{algorithmic}
\usepackage{graphicx}
\usepackage{textcomp}
\usepackage{xcolor}
\usepackage[numbers]{natbib}
\usepackage{siunitx}

\usepackage{booktabs}
\usepackage{diagbox} %

\usepackage[top=0.75in, left=0.75in, right=0.75in, bottom=0.75in]{geometry}
\def\IEEEtitletopspace{18pt}

\usepackage[normalem]{ulem} %

\usepackage{todonotes}
\def\BibTeX{{\rm B\kern-.05em{\sc i\kern-.025em b}\kern-.08em
    T\kern-.1667em\lower.7ex\hbox{E}\kern-.125emX}}

\begin{document}

\title{Real-Time Applicability of Emulated Virtual Circuits for Tokamak Plasma Shape Control\\
\thanks{This work was funded under the \textit{Fusion Computing Lab} collaboration between STFC Hartree Centre and UK Atomic Energy Authority.}
}

\author{
\IEEEauthorblockN{Pedro Cavestany}
\IEEEauthorblockA{\textit{STFC Hartree Centre} \\
\textit{SciTech Daresbury}\\
Warrington, \\
WA4 4AD, UK \\
pedro.cavestany-olivares@stfc.ac.uk
}
\and
\IEEEauthorblockN{Alasdair Ross}
\IEEEauthorblockA{\textit{STFC Hartree Centre} \\
\textit{SciTech Daresbury}\\
Warrington, \\
WA4 4AD, UK \\
alasdair.ross@stfc.ac.uk
}
\and
\IEEEauthorblockN{Adriano Agnello}
\IEEEauthorblockA{\textit{STFC Hartree Centre} \\
\textit{SciTech Daresbury}\\
Warrington, \\
WA4 4AD, UK \\
ORCID 0000-0001-9775-0331}
\and
\IEEEauthorblockN{Aran Garrod}
\IEEEauthorblockA{\textit{STFC Hartree Centre} \\
\textit{SciTech Daresbury}\\
Warrington, \\
WA4 4AD, UK \\
aran.garrod@stfc.ac.uk}
\and
\IEEEauthorblockN{Nicola C. Amorisco}
\IEEEauthorblockA{\textit{UK Atomic Energy Authority}\\
Culham Campus,
Abingdon, \\
Oxfordshire, \\
OX14 3DB, UK \\
nicola.amorisco@ukaea.uk}
\and
\IEEEauthorblockN{George K. Holt}
\IEEEauthorblockA{\textit{STFC Hartree Centre} \\
\textit{SciTech Daresbury}\\
Warrington, \\
WA4 4AD, UK \\
ORCID 0000-0001-6814-9117}
\and
\IEEEauthorblockN{Kamran Pentland}
\IEEEauthorblockA{\textit{UK Atomic Energy Authority}\\
Culham Campus,
Abingdon, \\
Oxfordshire, \\
OX14 3DB, UK \\
kamran.pentland@ukaea.uk
}
\and
\IEEEauthorblockN{James Buchanan}
\IEEEauthorblockA{\textit{UK Atomic Energy Authority}\\
Culham Campus,
Abingdon, \\
Oxfordshire, \\
OX14 3DB, UK \\
ORCID 0009-0009-0743-7655}
}

\maketitle

\begin{abstract}
Machine learning has recently been adopted to emulate sensitivity matrices for real-time magnetic control of tokamak plasmas. However, these approaches would benefit from a quantification of possible inaccuracies.
We report on two aspects of real-time applicability of emulators.
First, we quantify the agreement of target displacement from VCs computed via Jacobians of the shape emulators with those from finite differences Jacobians on exact Grad--Shafranov solutions.
Good agreement ($\approx$5-10\%) can be achieved {on a selection of geometric targets using} combinations of neural network emulators with $\approx10^5$ parameters. A sample of $\approx10^{5}-10^{6}$ synthetic equilibria is essential to train emulators that are not over-regularised or overfitting. Smaller models trained on the shape targets may be further fine-tuned to better fit the Jacobians.
{Second,} we address the effect of vessel currents that are not directly measured in real-time {and} are typically subsumed into effective \textit{``shaping currents''} {when designing} virtual circuits.
{We demonstrate that} shaping currents can be inferred via simple linear regression on a trailing window of active coil current measurements with residuals of only a few Amp{\`e}res, {enabling a} choice {for the} most appropriate shaping currents at any point {in a} shot.
While these results are based on historic shot data and simulations tailored to MAST-U, they indicate that emulators with few-millisecond latency can be developed for robust real-time {plasma shape} control in existing and upcoming tokamaks.
\end{abstract}

\section{Introduction}
One core task in magnetic confinement fusion with tokamaks is the control of the plasma position and shape, to maintain configurations with desirable confinement properties. This often amounts to the real-time control of a few geometric targets related to the  boundary of the plasma
\citep[under a toroidal symmetry approximation,][]{deTommasi2007,Ariola2008, Ambrosino2008}. The plasma is shaped by magnetic fields from
surrounding poloidal field (PF) coils, which must also compensate for the
``stray'' varying magnetic field from Ohmic coils that are used to drive a plasma
current. Typically, feed-forward sequences of desired target values are pre-computed, and
control loops are designed to correct for any departure of the plasma control
targets, on sub-millisecond timescales \citep{McArdle2020}. The
configuration of the plasma is established on faster (\qtyrange[range-units=single,range-phrase=-]{1}{10}{\micro\second}) timescales and can be described by equilibrium solutions that satisfy the Grad--Shafranov (GS) equation. For each equilibrium, a sensitivity
matrix can be computed as a Jacobian of the shape targets of interest with
respect to the PF coil currents. Its pseudo-inverse then provides a set of
\textit{virtual circuits} {(VCs)}, i.e. linear combinations of PF coil current changes
that would correspond to independent changes in one shape target at a time. This
is the approach commonly followed in classical shape control, where the plasma
control system issues voltage requests to the power supplies to attain desired
PF current changes to correct a departure between the desired and observed shape
targets. 

Approaches from machine learning have been recently proposed in the literature. Reinforcement learning (RL) can directly propose the voltage requests without
any linearisation assumption \citep{Seo2021,deGrave2022,Zhang2024}, and the RL agents are
trained on purely simulated environments or using
information from experimental campaigns. Compared to
classical control, RL has the advantage that resulting control policies can compensate for any effect that can be explicitly modelled in the training environment, e.g. currents induced in other metals in the vessel, which are not obvious to implement in classical control.
The generalisation and explainability of RL in plasma shape control, {however,} remain an open problem \citep{Tracey2024}. %

While the classical approach with virtual circuits only requires Jacobians to be computed around configurations of interest, different configurations correspond to different Jacobians and {typically} only a few are pre-computed for every experiment, hindering the robustness of control against non-negligible displacements.

Machine learning emulation can generalise the classical-control approach by providing target Jacobians at every time step of any foreseen experiment with very low latency.
{It also enables} better explainability than RL as the remainder of the control loops {are} driven by a small number of parameters that can be tuned by the operators.
Data-driven emulation of plasma shape has been developed for some tokamaks with long histories of campaigns \citep{Wai2022} and also for data-sets that are built entirely on synthetic libraries of GS solutions \citep{Agnello2024}. %

This work examines two aspects of emulated virtual circuits that affect their applicability in real-time plasma shape control. In Section~\ref{sec:jacobians}, we compare the Jacobians of emulated shape targets with those obtained from finite differences of exact GS solutions to quantify the accuracy of {the} emulated VCs, and the target displacements obtained by re-solving GS equilibria after the VCs are applied. In Section~\ref{sec:shaping}, we analyse the effect of vessel currents and provide procedures to account for their effect in shaping the plasma. 
{We use emulators trained on large synthetic libraries of equilibria, generated
using a Markov-Chain Monte Carlo method \citep{Agnello2024}, with the initial conditions seeded near real MAST-U \citep[Mega-Amp{\`e}re Spherical Toakamak Upgrade,][]{Fishpool2013} equilibria.}

\subsection{Notation}
\begin{figure}[htbp]
\centerline{
\includegraphics[width=0.85\columnwidth]{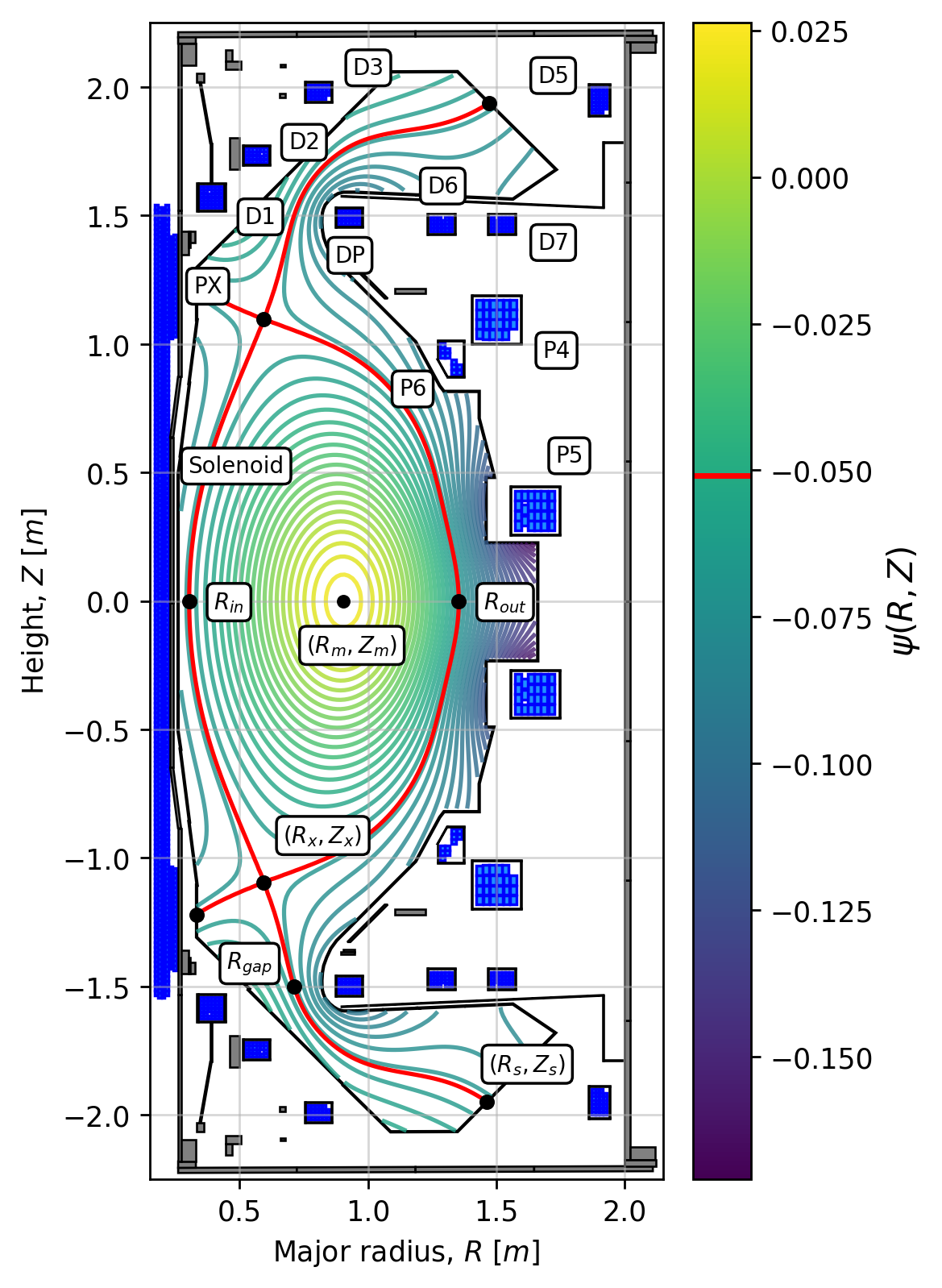}}
\vspace{-0.2in}
\caption{A poloidal cross-section of a plasma equilibrium from MAST-U \citep[shot $45456$ at $t=0.72$s, resolved with FreeGSNKE][]{Amorisco2024,Pentland2024}, delimited by the last closed flux surface (in red). Labels mark the basic shape control targets, PF and Ohmic coils are in blue and passive vessel structures are in gray. The contours of total $\psi$ are plotted only within the ``plasma domain'' bounded by the inner wall structures. %
}
\label{fig:equi_cross_section}
\end{figure}

This work is rather dense on specific notation from tokamak plasma physics, so we provide this short section as a reference. A collection of currents $\{I_{c}\}_{c=1,...,n_c}$ will also be denoted by a vector $\underline{I}_{(c)},$ where the subscript $(c)$ may indicate actively-controlled PF coils $(c=a)$ or passive vessel structures $(c=v).$ The regression slopes in Section~\ref{sec:shaping} are written as $b^{(c,\tau)}$ to distinguish them from the ``plasma beta'' ratios of thermodynamic and magnetic pressures. The MAST-U machine description and coil nomenclatures are provided by \cite{Fishpool2013,Agnello2024}. Following \citep{Freidberg2008}, throughout this paper the flux function $\psi(R,Z)$ is defined from the vertical magnetic field $B_{Z}$ as
\begin{equation}
    \psi(R,Z)=\int_{0}^{R}B_{Z}(R^{\prime},Z)R^{\prime}\mathrm{d}R^{\prime}\ 
\end{equation}
with radial and vertical coordinates measured from the tokamak axis and from the midplane respectively (see Figure~\ref{fig:equi_cross_section}).

\section{Validation of emulator Jacobians}
\label{sec:jacobians}

In this work, the input parameters considered for the emulators are the coil currents and plasma current density profile coefficients. The targets of interest are the inner and outer radii at the midplane ($R_\text{in},$ $R_\text{out}$), the coordinates of lower the X-point {($R_X, Z_X$)} , and the radial position of the lower divertor ``nose gap'' ($R_\text{gap}$) and strike-point {($R_{s}$)}, as shown in Figure~\ref{fig:equi_cross_section}.

\subsection{Data and emulators} In view of new machines, and to include
configurations not yet explored, large synthetic datasets {of
plasma equilibria} are necessary. A random combination of input parameters is very unlikely to
result in configurations with closed flux surfaces or desirable properties (e.g.
divertor leg strike points or lower X-point positions). {To achieve this,} we build
upon previous work \citep{Agnello2024}. Starting from reasonable parameter
choices, a Markov-Chain Monte Carlo (MCMC) random walk algorithm sequentially
samples the input parameters, {promoting} equilibrium configurations with desirable
properties. The random walk enables faster convergence of the GS solver, and the
parameter step sizes %
are adaptively adjusted.

We use FreeGSNKE \citep{Amorisco2024,Pentland2024}, to obtain GS equilibria for the chosen combinations of plasma parameters and PF coil currents.  
While \cite{Agnello2024} built entirely synthetic libraries, we start from snapshots of MAST-U experiments as initial seeds for each MCMC. 
We obtain $\approx1.8\times10^6$ equilibria with robustly computed targets and a separatrix entirely contained in the vessel, which include configurations encountered in MAST-U campaigns so far and extend to yet unseen configurations. From this, we separate a \textit{holdout set} of $3.5\times10^5$ equilibria close to the MAST-U MCMC seeds, which is never used to train or evaluate the shape emulators.

Following \cite{Agnello2024}, the emulators are fully-connected, feed-forward neural networks (FNNs), {because} they fulfil universal approximation theorems and are compositions of functions with defined and non-trivial gradients almost everywhere.
We train the FNNs to reproduce shape control targets versus PF coil currents, plasma current, internal inductance and poloidal beta \citep[see e.g.][for definitions]{Ariola2008,Freidberg2008}. The emulators are trained to minimise the mean-absolute deviation over a training set, plus an $L^{2}$ regularisation loss whose amplitude is a hyperparameter to be tuned \citep{Agnello2024}. We draw a random subset of the cleaned data minus the holdout, and separate a random 20\% test set. We split the remainder into an 80\% training and 20\% validating set whenever a choice of hyperparameters is made, and implement early stopping and learning-rate shrinkage by monitoring the FNN loss (mean absolute error) on the validation set. 
We examined both a Thompson sampler and an annealing sampler to tune the FNN hyperparameters. We run training and tuning experiments for different choices of dataset size, up to $10^6.$ Optimal models have $\approx$5-10 layers and 100-200 nodes per layer.
Aggregating the {five} best FNNs achieves 1\% relative residual scatter ($\sqrt{1-R^{2}}$) and $<$0.03\% relative bias in the target predictions. The results in this work are reported for a Leaky Rectified Linear Unit activation function, although we have also trained emulators with Exponential Linear Unit and sigmoid activations, trained on 1.6$\times10^{5}$ equilibria.

\begin{figure*}[htbp]
    \centerline{\includegraphics[width=0.99\textwidth]{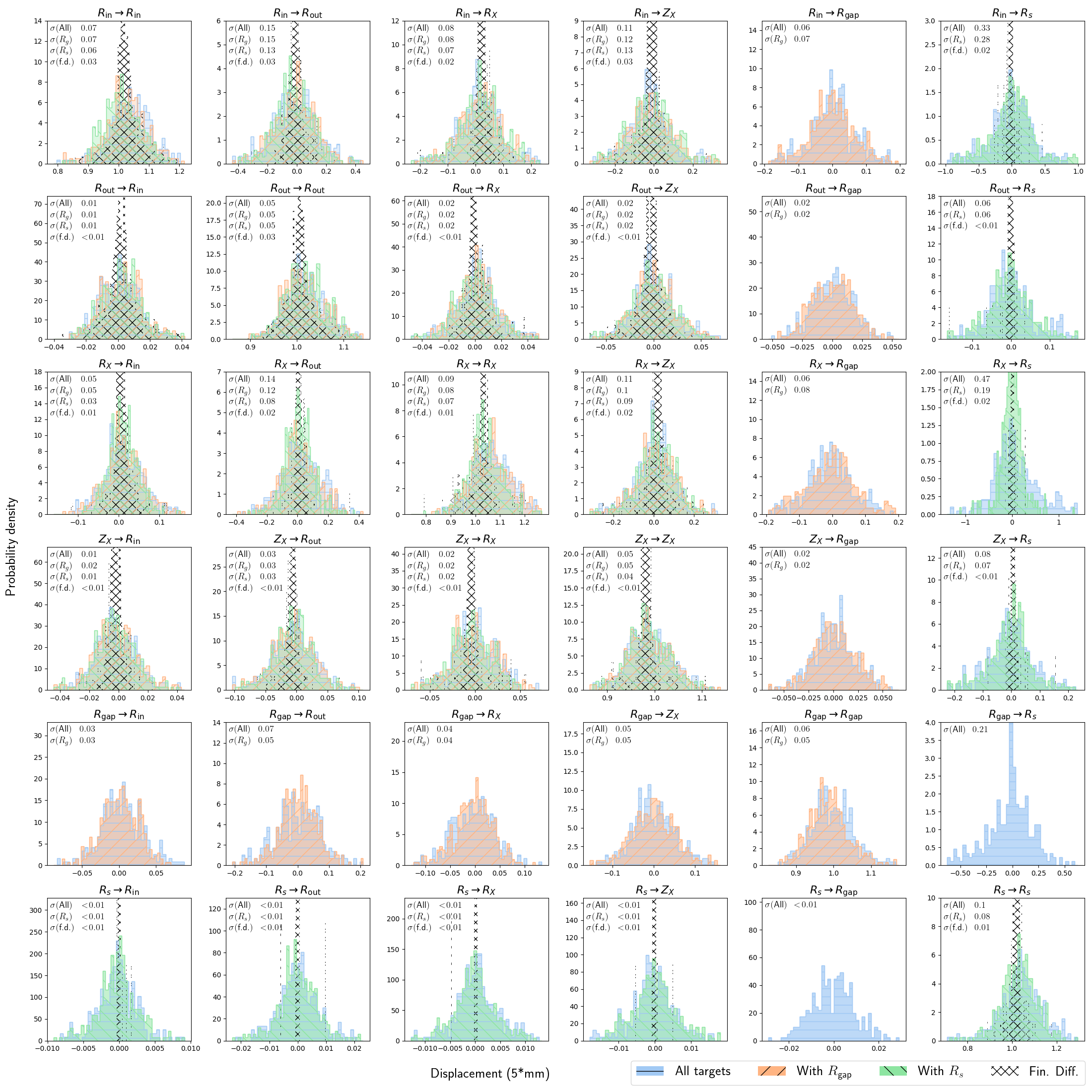} }
\caption{Displacements in shape targets ($R_\text{in}$, $R_\text{out}$, $R_{X}$,
    $Z_{X}$, $R_\text{gap}$, $R_{s}$) across $380$ equilibria in the holdout set after applying the VCs from the emulator-based Jacobians (coloured histograms) or the FreeGSNKE finite-difference Jacobians (black), followed by solving the GS equation.
All displacements are in units of the requested 5~mm shift in shape targets.
    Axis ranges are proportional to the standard deviations in the targets. The
    reported standard deviations are for VCs applied to the core shape targets
    ($R_\text{in},$ $R_\text{out},$ $R_{X},$ $Z_{X}$) combined with $R_\text{gap}$ or $R_{s}$ or both. 
}
\label{fig:displ_overl_emus_jtor}
\end{figure*}
\subsection{Jacobians and VC displacements}

While FNNs are universal approximations of $L^{p}$ functions, there is no guarantee that their derivatives {are also} a good fit {to the target function derivatives.} 
The FNNs are {trained on} the shape targets instead of their Jacobians, as a lower-dimensional target space {allows for simpler} FNN architectures and therefore more lenient {weight} regularisation. We quantify the accuracy of the emulators by computing Jacobians and their pseudo-inverse VCs around a random subset of 380 equilibria in the holdout set defined above. We use these VCs to request 5 mm displacements in each of the considered shape targets: we apply the relevant VC to the PF coil currents and resolve the GS problem to obtain the actual displacements. We compare such realised displacements with those resulting from application of the FreeGSNKE VCs, obtained via finite differences.

Each row in Figure~\ref{fig:displ_overl_emus_jtor} refers to a desired
displacement in each of the shape targets, and shows the histograms of the
actual displacements in all targets. Different coloured histograms correspond to
including different subsets of targets when computing the VCs. Perfect
decoupling of the targets by the VCs would correspond to a peak at 1 on the
diagonal panels and at 0 elsewhere. While the accuracy of the numerical VCs is
superior (typically percent level), the VCs from the emulators also display,
overall, good performances (typically $5-10\%$). This is especially the case on
the `core' shape targets ($R_\text{in},$ $R_\text{out},$ $R_{X},$ $Z_{X}$), where the
performance is often comparable to the finite difference VCs. When compared to
the finite difference VCs, the performance is poorer for requested shifts in
$R_{s}$, although the actual spread in that case is only at percent level.
Requesting VCs that can decouple both $R_\text{gap}$ and $R_{s}$ appears more
challenging and would warrant further exploration. It should also be noted that
calculation of `exact' finite difference VCs is not without its challenges. In
absence of auto-differentiable GS solvers, finite differences need to be
carefully sized to result in effective VCs. The results relating to finite
difference VCs presented here are computed by referring to a relative variation
of 0.002 in the plasma current density distribution. We have found that,
typically, finite difference VCs computed by referring to the relative variation
in the targets themselves, are less accurate.

\section{Shaping currents}
\label{sec:shaping}
Besides the PF coils, {whose currents are} actively controlled via the power supplies, other {toroidally symmetric} \textit{passive} conductors are present all around the plasma, e.g. vessel metals and graphite tiles in the divertor. 
As the PF currents and plasma current distribution change through a shot, they can induce significant currents in the passive structures which in turn {make} a non-negligible contribution to {$\psi$ and therefore the plasma shape.} %
Typically, \textit{shaping currents} are defined that seek to reabsorb the passive structure contribution, i.e. currents {that would be needed in} the PF coils (alone) to produce the same {flux $\psi$} as the full set of {currents in the} PF coils and passive structures. %
This approach is used when virtual circuits are pre-computed for new experiments using reconstructed equilibria from a previous shot. When such VCs are applied in the control loop, the requested currents {are in fact the} shaping currents.

\subsection{Offline inference of shaping currents}

\begin{figure*}[htbp]
    \centerline{\includegraphics[width=\textwidth]{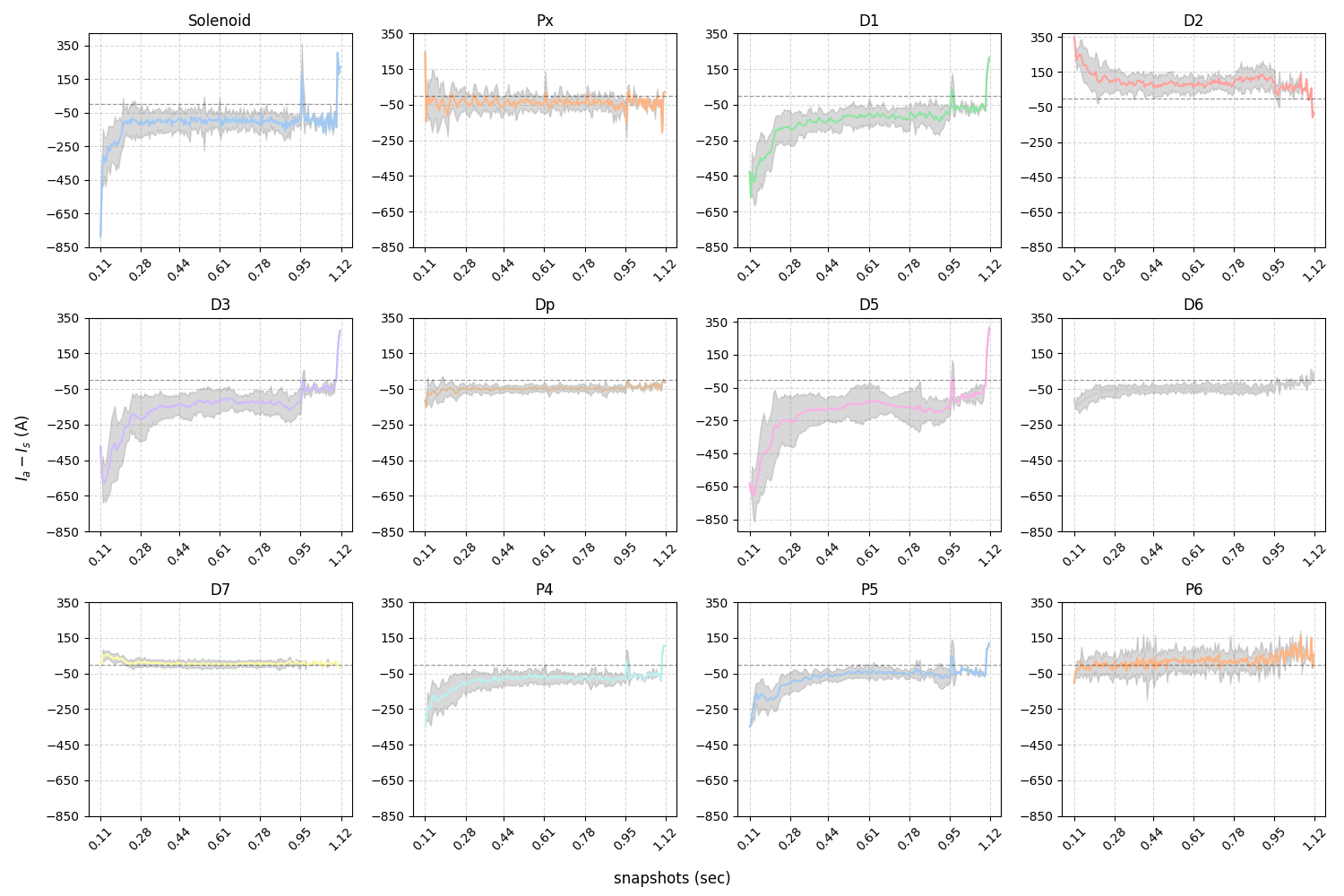}}
\caption{Differences between shaping currents and measured coil currents for 21 MAST-U shots. Each plot shows $\mu \pm \sigma$ of the difference across all shots for all snapshots in the corresponding active current. These differences can be fit with simple regressions that result in few-Amp{\`e}res residuals.
}
\vspace{-0.2in}
\label{fig:diff_Ieff_Ic}
\end{figure*}

{The} active PF coil currents $\left\{ I_{a}\right\}_{a=1,...,n_a}$ and passive vessel currents $\left\{ I_{v}\right\}_{v=1,...,n_v}$ produce the {\emph{tokamak flux}} %
\begin{equation}
\psi_{tok}(R,Z)=\sum_{c\in a\oplus v}\mathcal{G}(R,Z;R_{c},Z_{c})I_{c}.
\end{equation}

Over a computational grid of points $(R_i,Z_j)$, the Green functions $\mathcal{G}$ generate matrices of known coefficients $G_{i,j}^{c}=\mathcal{G}(R_{i},Z_{j};R_{c},Z_{c}),$ {so the} shaping currents can be inferred via ordinary linear regression 
\begin{equation}
    \underline{I}_{(s)}=(\mathbf{G}^{T}\mathbf{G})^{\dag}\mathbf{G}^{T}\underline{\psi}_{tok},
\end{equation}
where $\mathrm{dim} (\underline{I}_{(s)}) = n_a$ and the dagger denotes the pseudo-inverse. %
{This can be calculated, for each equilibrium in the dataset,} using only the active coil Green function matrix elements.
The regression is {carried out over} $(R_{i},Z_{j})$ points within {the limiter/wall only} %
(see caption of Fig.~\ref{fig:equi_cross_section}).

\begin{figure}[hbpt]
\centerline{\includegraphics[width=0.95\columnwidth]{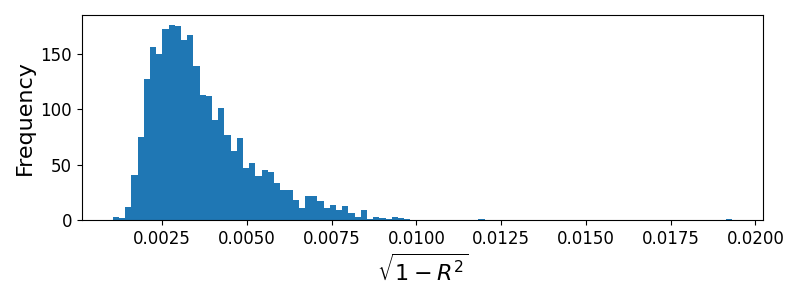}}
\centerline{\includegraphics[width=0.95\columnwidth]{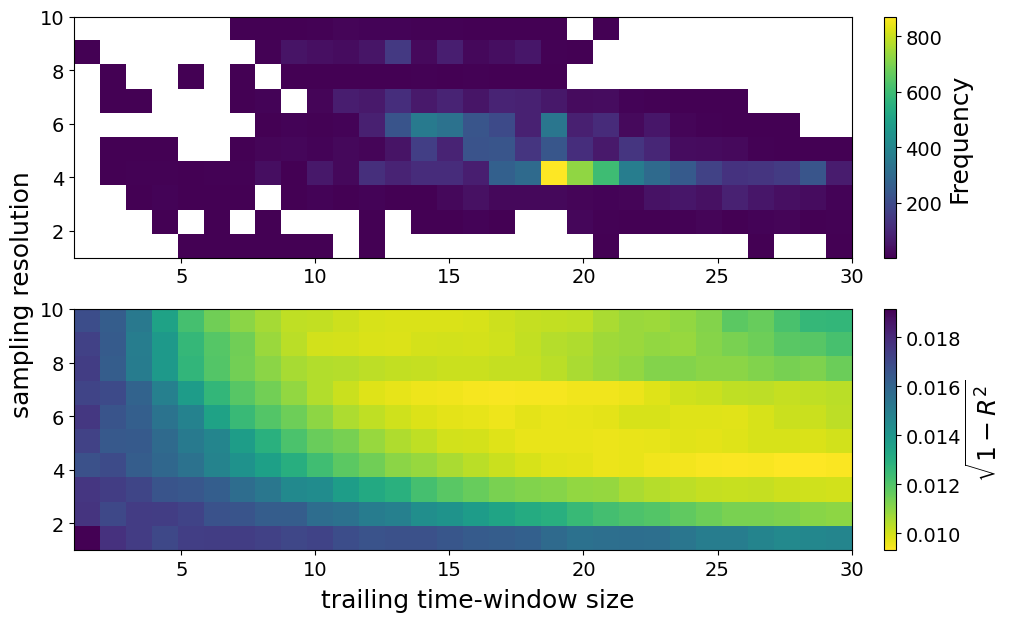} }
    \caption{Distribution of relative residual scatter in $\psi_{tok}(R_{i},Z_{j})$ from fits with shaping currents for 21 MAST-U shots.
    \textit{Middle:} Occurrences when each setup of regressors for the online inference (eq.~\ref{eq:Is_fit}) was the best solution, over $10^4$ runs with randomised train/test splits.
    \textit{Bottom:} Performance of the online inference of shaping currents as linear combinations across different setups for one of the random train/test splits.
    }
    \vspace{-0.2in}
\label{fig:psi_r2}
\end{figure}

The validation of this model is performed on \num{\approx 2700} snapshots across 21 {different} MAST-U shots. 
We restrict the inference and validation to {snapshots from} the ``flat-top'' phases (where plasma current is roughly constant and maximal for the discharge), since transient currents in passive structures tend to be stronger in the {ramp-up/down phases}, which are not usually controlled with active feedback loops.

The inferred shaping currents can differ from the PF coil currents with statistical significance across the 21 shots, as shown in Figure~\ref{fig:diff_Ieff_Ic}. Based on the $R^2$ metric, the shaping current model explains more than $98\%$ of variability of $\psi_{tok}$ in all validation cases, with a residual scatter $\sqrt{1 - R^2}$ below $0.01$ for the majority of validation snapshots, as shown in Figure~\ref{fig:psi_r2}. This result means that, at least for MAST-U shots, the equilibria can indeed be parameterised by a set of effective currents at the PF coil locations, albeit with different values than those measured in real time.

This ``offline inference'' requires a sufficiently accurate $\psi_{tok}$ reconstruction, which is not necessarily performed in real time. The {MAST-U} equilibria {in our dataset} have a sampling rate of 5~ms, whereas the active currents are measured every 0.6~ms.

\subsection{Online inference of shaping currents}

Due to the inductive coupling of active and passive PF structures, we may then seek a linear surrogate of the shaping currents using a trailing time-window of active current measurements in conjunction with the plasma current $I_{p}$. These are integrated quantities that can be measured or inferred with sub-millisecond cadence.
In particular, we examine linear regressions of the form
\begin{equation}
    I_{s}(t)-I_{a}(t)\sim \sum_{a^{\prime},\tau} b^{(a^{\prime},\tau)}I_{a^{\prime}}(t-\tau)+b^{(p,\tau)}I_{p}(t-\tau),
\label{eq:Is_fit}
\end{equation}
with $s,a=1,...,n_a,$ where $\tau$ denotes some chosen time-lag in a trailing window, and the matrix $ \mathbf{B} = [b^{(a,\tau)} \rvert b^{(p,\tau)}]$ is to be found. Here, $\mathbf{B}$ is posited to be the same across all shots, i.e. an inherent property of the tokamak. 

Different choices for the window {size} and sampling {resolution} of past readings (at $t-\tau$) can affect the quality of the inference. 
We train the regressions on 11 MAST-U shots, {using a different set of} 10 {more shots} as a validation set.
The results in Figure~\ref{fig:psi_r2} {show that} the optimal setup has a
sampling resolution of 4 time steps in the active currents and a trailing window of $\tau=19$ previous snapshots, corresponding to $\sqrt{1 - R^2}=0.015$ on the validation set. This setup was by far the most frequently chosen solution in a set of $10^4$ random train/test splits.

\section{Discussion}
Different methods from machine learning can provide fast and smooth policies for real-time control. Emulation of control targets can provide explainable and robust control policies, but like pre-computed lookup matrices its accuracy is limited by the approximations required for real-time applicability, in particular, that of sufficiently small models on reasonably low-dimensional input and target spaces.

The tests of Jacobian and VC displacement accuracy in Section~\ref{sec:jacobians} indicate that VCs appropriate to different equilibria can, in principle, be predicted by differentiable emulators trained on the shape targets (typical accuracy of 5-10\% in the resulting displacements). 
Training the emulators on the shape targets allows for more compact architectures that are less prone to overfitting and can provide low enough latency (a few ms for $10^{5}-10^{6}$ parameters). If higher accuracies are deemed necessary, the emulator models may be fine-tuned on the finite difference VCs themselves.

{During} feedback control, the voltage requests on power supplies are proportional to the {difference} between desired and measured shape targets -- as well as their numerical derivatives and integrals over a trailing time window.
{These differences are} parametrised in terms of currents because they are directly related to voltages through the circuit equations and can be more intuitively constrained to remain within operational limits. 
Currents induced in the vessel can be non-negligible, so the PF current corrections are better interpreted in terms of {the above-defined} shaping currents, which can best reproduce the same flux as the full set of PF coils and passive structures. 
The shaping currents can be computed offline, from post-experiment equilibrium reconstructions, {or as} linear combinations of PF coil currents over a sliding time window, with percent-level {accuracy} and statistically negligible residuals. 
Since the PF coil currents can be measured with sub-millisecond cadence, the shaping currents can also be computed and adjusted every millisecond. 
This is a substantial improvement over {existing approaches that use} virtual circuit lookup tables pre-computed for phases of the experiment that are separated by hundreds of milliseconds.


\begin{thebibliography}{00}
\bibitem{deTommasi2007} de Tommasi, G., et al., “XSC Tools: A Software Suite for Tokamak Plasma Shape Control Design and Validation”, \textit{IEEE Transactions on Plasma Science}, vol. 35, no. 3, IEEE, pp. 709–723, 2007. doi:10.1109/TPS.2007.896989.
\bibitem{Ariola2008} Ariola, M., et al., “Integrated Plasma Shape and Boundary Flux Control on JET Tokamak”, \textit{Fusion Science and Technology}, vol. 53, no. 3, pp. 789–805, 2008. doi:10.13182/FST08-A1735.
\bibitem{Ambrosino2008} Ambrosino, G., et al., “Plasma Strike-Point Sweeping on JET Tokamak With the eXtreme Shape Controller”, \textit{IEEE Transactions on Plasma Science}, vol. 36, no. 3, IEEE, pp. 834–840, 2008. doi:10.1109/TPS.2008.922920.
\bibitem{Ariola1999} Ariola, M., Ambrosino, G., Lister, J. B., Pironti, A., Villone, F., and Vyas, P., “A Modern Plasma Controller Tested on the TCV Tokamak”, \textit{Fusion Technology}, vol. 36, no. 2, pp. 126–138, 1999. doi:10.13182/FST99-A97.
\bibitem{McArdle2020} McArdle, G., Pangione, L., and Kochan, M., “The MAST Upgrade plasma control system”, \textit{Fusion Engineering and Design}, vol. 159, Art. no. 111764, 2020. doi:10.1016/j.fusengdes.2020.111764.
\bibitem{Seo2021}Seo, J., “Feedforward beta control in the KSTAR tokamak by deep reinforcement learning”, \textit{Nuclear Fusion}, vol. 61, no. 10, Art. no. 106010, IOP, 2021. doi:10.1088/1741-4326/ac121b.
\bibitem{deGrave2022} Degrave, J., et al., “Magnetic control of tokamak plasmas through deep reinforcement learning”, \textit{Nature}, vol. 602, no. 7897, pp. 414–419, 2022. doi:10.1038/s41586-021-04301-9.
\bibitem{Zhang2024} Zhang, Y. C., Wang, S., Yuan, Q. P., Xiao, B. J., and Huang, Y., “Real-time feedback control of $\beta_{p}$ based on deep reinforcement learning on EAST”, \textit{Plasma Physics and Controlled Fusion}, vol. 66, no. 5, Art. no. 055014, IOP, 2024. doi:10.1088/1361-6587/ad3749.
\bibitem{Tracey2024} Tracey, B. D., et al., “Towards practical reinforcement learning for tokamak magnetic control”, \textit{Fusion Engineering and Design}, vol. 200, Art. no. 114161, 2024. doi:10.1016/j.fusengdes.2024.114161.

%

\bibitem{Wai2022} Wai, J. T., Boyer, M. D., and Kolemen, E., “Neural net modeling of equilibria in NSTX-U”, \textit{Nuclear Fusion}, vol. 62, no. 8, Art. no. 086042, IOP, 2022. doi:10.1088/1741-4326/ac77e6.

\bibitem{Agnello2024} Agnello, A., et al., “Emulation techniques for scenario and classical control design of tokamak plasmas”, \textit{Physics of Plasmas}, vol. 31, no. 4, Art. no. 043901, AIP, 2024. doi:10.1063/5.0187822.

\bibitem{Fishpool2013} Fishpool, G., et al., “MAST-upgrade divertor facility and assessing performance of long-legged divertors”, \textit{Journal of Nuclear Materials}, vol. 438, pp. S356–S359, 2013. doi:10.1016/j.jnucmat.2013.01.067.

\bibitem{Freidberg2008} Freidberg, J. P., \textit{Plasma Physics and Fusion Energy}, Cambridge, UK: Cambridge University Press, 2008.

\bibitem{Amorisco2024} Amorisco, N. C., Agnello, A., Holt, G., Mars, M., Buchanan, J., and Pamela, S., “FreeGSNKE: A Python-based dynamic free-boundary toroidal plasma equilibrium solver”, \textit{Physics of Plasmas}, vol. 31, no. 4, Art. no. 042517, AIP, 2024. doi:10.1063/5.0188467.

\bibitem{Pentland2024} Pentland, K., et al., “Validation of the static forward Grad-Shafranov equilibrium solvers in FreeGSNKE and Fiesta using EFIT++ reconstructions from MAST-U”, \textit{Physica Scripta}, vol. 100, no. 2, Art. no. 025608, IOP, 2025. doi:10.1088/1402-4896/ada192.

\end{thebibliography}
\end{document}